\documentclass[11pt]{article} %
\usepackage{graphicx}
\usepackage{fancyhdr}
\usepackage{makeidx}
\usepackage{amssymb,amsmath,pdfpages}
\usepackage{physics}
\usepackage{gensymb}
\usepackage{upquote}
\usepackage{ulem}
\usepackage{xcolor}
\usepackage{cite}
\usepackage[utf8]{inputenc}
\usepackage[nottoc]{tocbibind}
\usepackage{setspace}
\usepackage{enumerate}
\DeclareUnicodeCharacter{0308}{HERE!HERE!}
\DeclareUnicodeCharacter{030C}{HERE!HERE!}
\DeclareUnicodeCharacter{0301}{HERE!HERE!}
\newcommand{\bd}{\begin{document}}
	\newcommand{\ed}{\end{document}}
\newcommand{\bc}{\begin{center}}
	\newcommand{\ec}{\end{center}}
\newcommand{\vs}{\vspace}
\newcommand{\hs}{\hspace}
\newcommand{\beq}{\begin{equation}}
\newcommand{\eeq}{\end{equation}}
\newcommand{\beqs}{\begin{eqn*}}
	\newcommand{\eeqs}{\end{eqn*}}
\newcommand{\bq}{\begin{quote}}
	\newcommand{\eq}{\end{quote}}
\newcommand{\lb}{\linebreak}
\newcommand{\mb}{\makebox}
\newcommand{\fb}{\framebox}
\newcommand{\mc}{\multicolumn}
\newcommand{\ben}{\begin{enumerate}}
	\newcommand{\een}{\end{enumerate}}
\newcommand{\bit}{\begin{itemize}}
	\newcommand{\eit}{\end{itemize}}
\newcommand{\ov}{\overline}
\newcommand{\un}{\underline}
\newcommand{\lt}{\left}
\newcommand{\rt}{\right}
\newcommand{\ba}{\begin{array}}
	\newcommand{\ea}{\end{array}}
\newcommand{\beqa}{\begin{eqnarray}}
\newcommand{\eeqa}{\end{eqnarray}}
\newcommand{\beqas}{\begin{eqnarray*}}
	\newcommand{\eeqas}{\end{eqnarray*}}
\newcommand{\bfg}{\begin{figure}}
	\newcommand{\efg}{\end{figure}}
\newcommand{\pad}{\partial}
\newcommand{\nn}{\nonumber}
\newcommand{\la}{\leftarrow}
\newcommand{\ra}{\rightarrow}
\newcommand{\lgla}{\longleftarrow}
\newcommand{\lgra}{\longrightarrow}
\newcommand{\La}{\Leftarrow}
\newcommand{\Ra}{\Rightarrow}
\newcommand{\Lra}{\Leftrightarrow}
\newcommand{\Lgla}{\Longleftarrow}
\newcommand{\Lgra}{\Longrightarrow}
\renewcommand{\a}{\alpha}
\renewcommand{\b}{\beta}
\newcommand{\g}{\gamma}
\newcommand{\G}{\Gamma}
\renewcommand{\d}{\delta}
\newcommand{\D}{\Delta}
\newcommand{\e}{\epsilon}
\newcommand{\eps}{\epsilon}
\newcommand{\s}{\sigma}
\renewcommand{\l}{\lamda}
\newcommand{\m}{\mu}
\newcommand{\n}{\nu}
\renewcommand{\S}{\Sigma}
\newcommand{\p}{\pi}
\newcommand{\om}{\omega}
\newcommand{\Om}{\Omega}
\newcommand{\tri}{\triangle}
\newcommand{\ti}{\times}
\newcommand{\f}{\frac}
\newcommand{\ds}{\displaystyle}
\newcommand{\bm}[1]{\mb{{\boldmath $#1$}}}
\newcommand{\alter}[2]{\lt\{ \ba{ll}#1 \\ #2 \ea \rt.}
\newcommand{\alt}[4]{\lt\{ \ba{ll}#1 & \mb{if \, \,}#2 \\ #3 & \mb{}#4 \ea
	\rt.}
\newcommand{\altn}[4]{\lt\{ \ba{rl}#1 & \mb{if \, \,}#2 \\ #3 & \mb{}#4 \ea
	\rt.}
\newcommand{\altif}[4]{\lt\{ \ba{ll}#1 & \mb{if \, \,}#2 \\ #3 &
	\mb{if \, \,}#4 \ea \rt.}
\newcommand{\altnif}[4]{\lt\{ \ba{rl}#1 & \mb{if \, \,}#2 \\ #3 &
	\mb{if \, \,}#4 \ea \rt.}
\newcounter{algc}
\newcounter{romc}
\newcounter{Alphc}
\newcommand{\bl}{\begin{list}{{\it Step} ~\arabic{algc}~:} {\usecounter{algc}
			\setlength{\topsep}{0pt} \setlength{\itemsep}{0pt}}}
	\newcommand{\el}{\end{list}}
\newcommand{\blr}{\begin{list}{~\roman{romc}~:} {\usecounter{romc}
			\setlength{\topsep}{0pt} \setlength{\itemsep}{0pt}}}
	\newcommand{\elr}{\end{list}}
\newcommand{\bla}{\begin{list}{~\Alph{Alphc}~:} {\usecounter{Alphc}
			\setlength{\topsep}{0pt} \setlength{\itemsep}{0pt}}}
	\newcommand{\ela}{\end{list}}
\newcommand{\tsup}{\textsuperscript}
\newcommand{\tsub}{\textsubscript}

\newtheorem{theorem}{Theorem}
\begin{document}
\title{Harmonic to anharmonic tuning of moir\'e potential leading to unconventional Stark effect and giant dipolar repulsion in WS$_2$/WSe$_2$ heterobilayer}
\author{Suman Chatterjee$^{1,||}$, Medha Dandu$^{1,2,||}$, Pushkar Dasika$^{1,||}$, Rabindra Biswas$^{1}$, \\Sarthak Das$^{1,3}$, Kenji Watanabe$^4$, Takashi Taniguchi$^5$,\\ Varun Raghunathan$^1$, and Kausik Majumdar$^{1*}$\\
	$^1$Department of Electrical Communication Engineering, \\Indian Institute of Science, Bangalore 560012, India\\	
	$^2$Currently with Molecular Foundry, Lawrence Berkeley National Laboratory,\\  Berkeley, CA 94720, United States\\
	$^3$Currently with Institute of Materials Research and Engineering (IMRE),\\ Agency for Science, Technology and Research (A*STAR),\\ Singapore 138634, Republic of Singapore\\
    $^4$Research Center for Functional Materials,\\ National Institute for Materials Science, 1-1 Namiki, Tsukuba 305-044, Japan\\
	$^5$International Center for Materials Nanoarchitectonics,\\ National Institute for Materials Science, 1-1 Namiki, Tsukuba 305-044, Japan\\
	\\$^{||}$These authors contributed equally\\
	$^*$Corresponding author, email: kausikm@iisc.ac.in}
\date{}
\maketitle
\newpage
\begin{abstract}
Excitonic states trapped in harmonic moir\'e wells of twisted heterobilayers is an intriguing testbed. However, the moir\'e potential is primarily governed by the twist angle, and its dynamic tuning remains a challenge. Here we demonstrate anharmonic tuning of moir\'e potential in a WS$_2$/WSe$_2$ heterobilayer through gate voltage and optical power. A gate voltage can result in a local in-plane perturbing field with odd parity around the high-symmetry points. This allows us to simultaneously observe the first (linear) and second (parabolic) order Stark shift for the ground state and first excited state, respectively, of the moir\'e trapped exciton - an effect opposite to conventional quantum-confined Stark shift. Depending on the degree of confinement, these excitons exhibit up to twenty-fold gate-tunability in the lifetime ($100$ to $5$ ns). Also, exciton localization dependent dipolar repulsion leads to an optical power-induced blueshift of $\sim$1 meV/$\mu$W - a five-fold enhancement over previous reports.
  \end{abstract}

\section*{Introduction}
Interlayer van der Waals interaction allows us to stack layers of transition metal dichalcogenides (TMDCs) onto each other with an arbitrary lattice mismatch \cite{chiu2015determination,cheng2014electroluminescence,dandu2022electrically}. This leads to an additional degree of freedom, the twist angle ($\theta$) between two successive layers, that governs the moir\'e pattern arising in the corresponding superlattice \cite{tran2019evidence,mak2022semiconductor,wu2018theory,lau2022reproducibility}. The lattice constant of the moir\'e is given by $a_M \approx \frac{a}{\sqrt{\theta^2+\delta^2}}$ where $\delta$ is the lattice constant difference and $a$ being the average lattice constant  \cite{yuan2020twist,jin2019identification,wu2018theory}. Different atomic registries present in this moir\'e superlattice (Figure \ref{fig:introduction to three moire ILEs}a) form a periodic potential fluctuation [$V_M(\mathbf{r})$] resulting from local strain and interlayer coupling \cite{naik2020origin,naik2018ultraflatbands}. Varying twist angle can dramatically change the material properties, drawing attention from the researchers in the recent past \cite{lin2023room,chuang2022emergent,shi2019twisted,yuan2020twist}. Moir\'e superlattice in TMDC heterobilayer has been widely explored including observation of neutral moir\'e exciton \cite{tran2019evidence,alexeev2019resonantly,seyler2019signatures}, moir\'e trion \cite{liu2021signatures,wang2021moire,marcellina2021evidence}, single photon emission \cite{mukherjee2020observation,kremser2020discrete}, and correlated states \cite{xu2020correlated,mak2022semiconductor,liu2021excitonic}.

Due to type-II band alignment, WS\tsub2/WSe\tsub2 heterobilayer supports an ultrafast charge transfer \cite{jin2018ultrafast,hong2014ultrafast} with electrons staying in the WS\tsub{2} conduction band, and holes in the WSe\tsub2 valance band, forming interlayer exciton (ILE) \cite{yuan2020twist,jin2019identification} under optical excitation (Figure \ref{fig:introduction to three moire ILEs}b). The moir\'e wells behave as two-dimensional harmonic traps for the ILEs \cite{tran2019evidence,tan2022signature,lohof2023confined}.

The depth of the exciton moir\'e potential is determined by the twist angle and the degree of lattice mismatch between the two heterobilayers. Hence, dynamic tuning of moir\'e potential remains a challenge, which, if realised, will be of great importance for both scientific exploration and applications. One could perturb the moir\'e potential by external stimulus, however, the perturbing potential may not necessarily be harmonic, breaking down the usual harmonic potential approximation for moir\'e well. In this work, we explore two such anharmonic perturbations to the WS$_2$/WSe$_2$ moir\'e potential well: the first one through a gate voltage which introduces anharmonic perturbation through screening at high doping regime; and the second one is through optical excitation which introduces the perturbing potential through ILE dipolar repulsion. In both cases, the harmonic to anharmonic switching of the moir\'e potential manifests through a corresponding change from an equal to unequal inter-excitonic spectral separation. In such a scenario, we explore several intriguing features of the moir\'e excitons, including giant lifetime tunability, anomalous Stark shift, and dipolar repulsion induced large spectral blueshift.

\section*{Results and Discussion}
\begin{figure}[!hbt]
	\centering
	\vs{-0.1in}
	\hs{-0.0in}
	\includegraphics[scale=0.46]{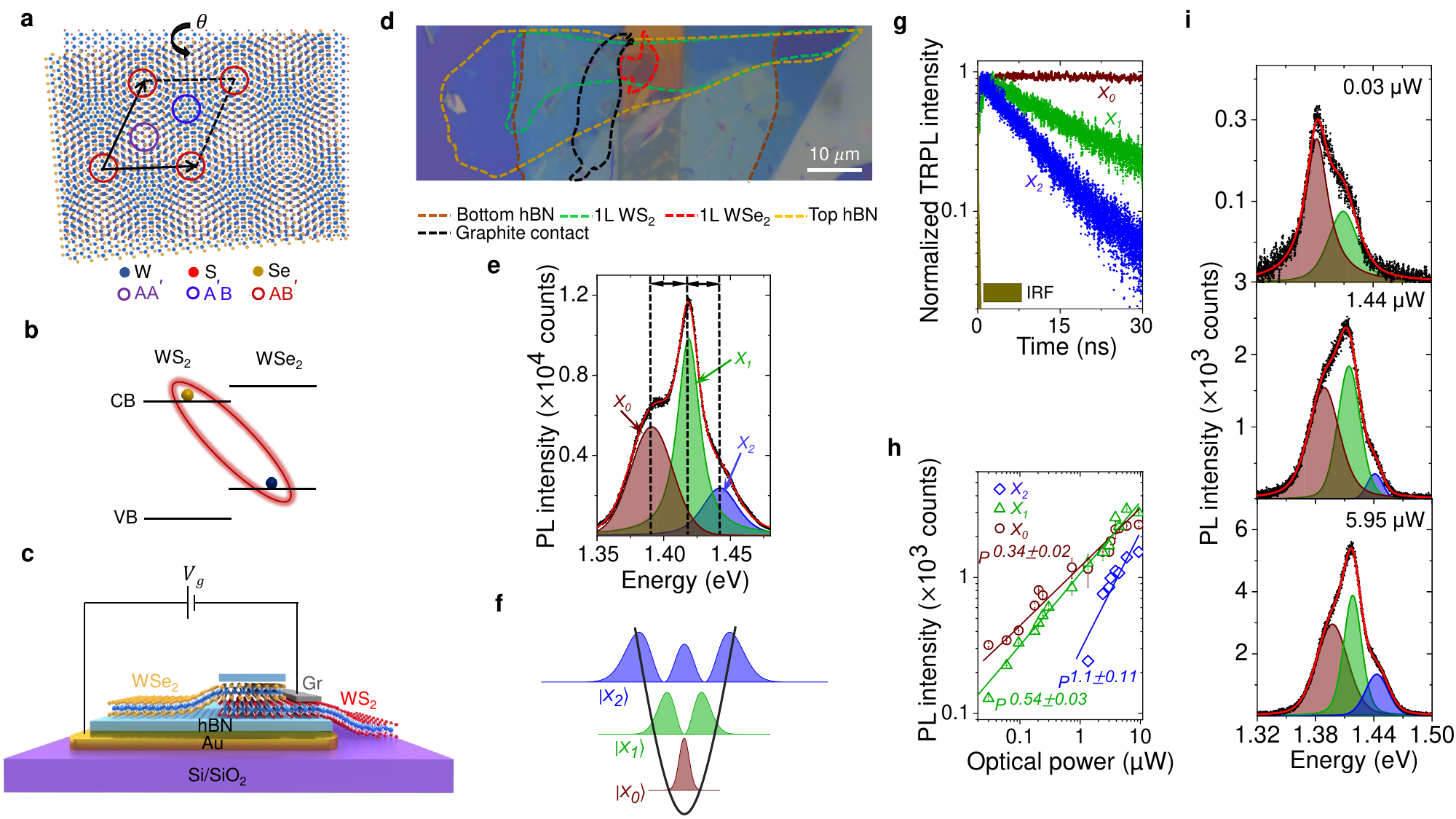}
	\vspace{-0.1in}
	\caption{\textbf{Moir\'e trapped interlayer exciton.} (a) Different atomic registries in a twisted WS$_2$/WSe$_2$ bilayer with high symmetry points marked by colored circles. (b) Type-II heterojunction of WS\tsub2/WSe\tsub2 bilayer resulting in interlayer exciton. (c)  Schematic of the heterobilayer with back gate connection. (d) Optical image of a fabricated device. The dotted colored lines indicate different flake boundaries. Scale bar is 10 $\mu$m. (e) Representative PL spectrum (using 532 nm CW laser) in the ILE regime (black symbols) and fitting (red trace) showing three clear ILE resonances denoted by $X_0$ (brown), $X_1$ (green), and $X_2$ (blue) at $V_g= 0$ V and $P=0.675$ $\mu$W. Black arrows indicate near equal spacing. (f) Schematic representation of three ILE states in a harmonic moir\'e potential well with varying degree of localization. (g) Raw TRPL spectra along with IRF for the three ILE resonances showing varying decay time scales at $V_g = 0$ V ($P=13.45$ $\mu$W), namely 100, 15, and 9.3 ns for $X_0$, $X_1$, and $X_2$, respectively. (h) Optical power dependent intensity plot (symbols) of the three ILEs in log-log scale following different power-laws (fitted by solid lines). (i) Evolution of power dependent PL spectra (black symbols) at three different optical powers, along with fitting (red solid line).} \label{fig:introduction to three moire ILEs}
\end{figure}

We prepare hBN-capped $\sim 59^\circ$ twisted (confirmed by second harmonic generation (SHG) spectroscopy in \textbf{Supplementary Note 1 and Figure 1}) WS\tsub2/WSe\tsub2 heterobilayer (sample D1) with a back gate (see \textbf{Methods} for sample preparation). The schematic and the optical image of sample D1 are illustrated in Figure \ref{fig:introduction to three moire ILEs}c and d. This twist angle creates a moir\'e superlattice with a lattice constant $\sim 7.3$ nm. Figure \ref{fig:introduction to three moire ILEs}e shows a representative photoluminescence (PL) spectrum from the sample with 532 nm excitation at 4 K. The emission spectrum exhibits three separate, strong interlayer moir\'e excitonic resonances \cite{sun2022enhanced} $X_0$, $X_1$, and $X_2$ at $\approx$ 1.392, 1.418 and 1.442 eV, respectively (marked by black dashed line). The peaks exhibit alternating sign of the degree of circular polarization (DOCP) (\textbf{Supplementary Figure 2}), indicating the existence of moir\'e superlattice \cite{tran2019evidence,yu2018brightened,wu2018theory}.

The near-equal inter-excitonic separation suggests that the three exciton resonances appear from excitonic states in the harmonic moir\'e potential well (Figure \ref{fig:introduction to three moire ILEs}f) \cite{tran2019evidence,tan2022signature,wu2018theory,lohof2023confined}. This inter-excitonic separation can be tuned by varying the twist angle, which regulates the depth of the moir\'e potential well \cite{tran2019evidence, choi2021twist}. We verified this by measuring twist angle dependent PL spectra from three samples [D1 ($\sim 59^\circ$), D2 ($\sim 54^\circ$) and D3 (large angle misalignment)] in \textbf{Supplementary Figure 3}. The time-resolved PL (TRPL) measurement (see \textbf{Methods}) from sample D1 in Figure \ref{fig:introduction to three moire ILEs}g shows that the lifetime of the three species ($\tau_{X_0}=100$ ns, $\tau_{X_1}=15.3$ ns, and $\tau_{X_2}=9$ ns) increases significantly with stronger confinement. Accordingly, their PL intensity also exhibits significantly different power law with varying optical power ($P$): $I \propto P^{\alpha_i}$ with $\alpha_0=0.34\pm0.02$, $\alpha_1=0.59\pm0.03$, and $\alpha_2=1.1\pm0.11$ (Figure \ref{fig:introduction to three moire ILEs}h). The corresponding spectra at three different $P$ values are shown in Figure \ref{fig:introduction to three moire ILEs}i. At low power ($30$ nW), $X_0$ emission is the dominant one, with negligible emission from $X_2$. However, at higher power ($5.95$ $\mu$W), three peaks are clearly discernable, and the fractional contribution of $X_0$ reduces, while $X_2$ emission becomes appreciable. All these observations indicate that the three different excitonic species correspond to moir\'e trapped excitonic states with varying degrees of localization (Figure \ref{fig:introduction to three moire ILEs}f). From the spectral separation between the quantized states, we calculate peak-to-peak moir\'e potential fluctuation of $\approx 150$ meV (see \textbf{Supplementary Note 2}), as shown in Figure \ref{fig:gate dependent ILE}c. Possible alternative explanations, such as phonon-sidebands and defect-bound excitons, are unlikely in our samples based on the observations including alternating signs of the DOCP and systematic tuning of the ILE peak separation with twist angle, doping, and optical power (discussed later).

\textbf{Gate tunability:} Figure \ref{fig:gate dependent ILE}a shows a color plot of the interlayer exciton emission spectra as a function of gate voltage ($V_g$). The estimated n-doping density at the highest applied $V_g$ ($=5$ V) is $<1.5 \times10^{12}$ cm$^{-2}$ (see \textbf{Supplementary Figure 4}). This is well below the moir\'e trap density ($n_0) \approx 2\times10^{12}$ cm$^{-2}$ for $a_M \sim 7.3$ nm. The fitted peak positions are shown in the left panel of Figure \ref{fig:gate dependent ILE}b (see individual spectra in \textbf{Supplementary Figure 5}). While the $V_g<$ 0 V region is nearly featureless, $V_g>$ 0 V (n-doping) region has three conspicuous features: (a) there is a reduction in emission intensity for all the three ILE peaks, with $X_0$ disappearing at high $V_g$; (b) there is a large and unequal redshift for the peaks for $V_g>0$; and (c) the inter-excitonic separation changes at higher $V_g$, indicating induced anharmonicity. The reduction in emission intensity with an increase in $V_g$ rules out the charged excitonic (trion) nature of any of the three peaks. Figure \ref{fig:gate dependent ILE}b (right panel) schematically explains the origin of the strong redshift with $V_g$. At positive $V_g$, the WS$_2$ layer becomes n-doped. Due to small thermal energy at 4 K, the wave function of the induced electrons remains primarily in the WS$_2$ layer, with a fraction of it extends into the WSe$_2$ bandgap as an evanescent state with imaginary wave vector. Such a wave function distribution creates a screening of the gate field, and in turn a relative potential difference between WS$_2$ and WSe$_2$ layers, reducing the interlayer bandgap. Note that, the presence of the charge density from the evanescent state in WSe$_2$ is essential to create such relative potential difference between the two layers, else dictated by the self-consistent electrostatics, a zero induced charge density in WSe$_2$ layer would result in pinning of the WSe$_2$ potential with that of WS$_2$, and no relative interlayer bandgap change would be allowed.

\textbf{Unconventional Stark effect:} Interestingly, the average slope (indicated by black dashed line in Figure \ref{fig:gate dependent ILE}b) of the redshift of $X_2$ is almost similar (about 5 meV/V) to that of the intra-layer WS$_2$ trion (X$^-$) or charged (XX$^-$) biexciton \cite{chatterjee2022trion} (See \textbf{Supplementary Figure 6}), but the average slope is higher for $X_1$ ($\sim$ 7 meV/V) and $X_0$ ($\sim$ 15 meV/V). The redshift of the intra-layer WS$_2$ trion emission peak with $V_g$ is directly related to the enhanced trion dissociation energy due to the extra energy required to place the remaining electron into the increasingly filled conduction band. Hence it can be correlated with the change in the Fermi energy due to doping \cite{chatterjee2022trion,mak2013tightly,kallatt2019interlayer}. This change is nearly equal to the shift in the WS$_2$ conduction band with respect to the WSe$_2$ valence band, making the average slopes of $X_2$ and WS$_2$ trion shift similar. This also is in agreement with the weak confinement of $X_2$.

However, the enhancement in the slope of the redshift for $X_1$ and $X_0$ cannot be explained from doping dependent interlayer bandgap reduction and suggests a strong additional effect of localization. To understand this further, we solve the 1D Poisson equation to obtain the movement of bands with $V_g$ (see \textbf{Supplementary Note 3} for the details of the calculation). The results are summarized in Figure \ref{fig:gate dependent ILE}d. At small positive $V_g$, the bands shift downward in energy (middle panel, $V_g=$ 0.5 V). However, at larger positive $V_g$, the central part of region I (right panel, $V_g>$ 0.5 V) of the conduction band moir\'e well being energetically closer to the Fermi energy supports more electron density than region II. Accordingly, due to the screening by the induced carrier density, region I starts moving down slower than region II. The net effect is a suppression in the local moir\'e fluctuation of the conduction band. Interestingly, the self-consistent electrostatics forces an amplification in the moir\'e potential fluctuation in the valence band of WSe$_2$: The suppressed movement of WS$_2$ bands in region I also reduces the movement of bands in WSe$_2$, while the stronger movement of WS$_2$ bands in region II (with relatively less carrier density) also pushes the WSe$_2$ bands more downward. The net result is a flattening of the electron moir\'e well in the WS\tsub2 conduction band, causing a delocalization of the electron state, coupled with a deeper hole moir\'e well in the WSe$_2$ valence band, resulting in an enhanced localization of the hole state (zoomed in Figure \ref{fig:gate dependent ILE}d, bottom panel). This modification of the moir\'e trapping potential, in turn, causes a reduction in the energy of the trapped electron state and an enhancement in the energy of the trapped hole state. The negative net change gives rise to an additional redshift in the localized exciton resonance ($X_0$ and $X_1$).

This results in an in-plane perturbation potential ($\Delta V$) with even parity about the high-symmetry points (Figure \ref{fig:gate dependent ILE}e). $\Delta V$ is maximum at the center of the moir\'e well and reduces symmetrically away from the center. On the other hand, the wave function ($\psi$) has an even and odd parity for the ground ($X_0$) and first excited ($X_1$) states, respectively. This, in turn, results in a large (small) value of $|\psi_0|^2$ ($|\psi_1|^2$) around the center of the trap for $X_0$ ($X_1$), as shown in Figure \ref{fig:gate dependent ILE}e. Due to such a strong overlap (non-overlap) of $\Delta V$ and $|\psi_0|^2$ ($|\psi_1|^2$), the first order Stark effect ($\bra{\psi}\Delta V\ket{\psi}$) is nonzero (negligible) for $X_0$ ($X_1$). Accordingly, we expect $X_0$ and $X_1$ to exhibit linear and parabolic Stark shift, respectively, with the in-plane local electric field ($\xi$), and hence with $V_g$, since our simulation suggests that $\xi$ is approximately linearly dependent on $V_g$ (see \textbf{Supplementary Figure 7}). Such local field effect will cancel out for the less-localized $X_2$ state. In Figure \ref{fig:gate dependent ILE}f, the respective Stark shifts [$\delta_{X_{0,1}}(V_g)-\delta_{X_{0,1}}(V_g=0)$ where $\delta_{X_0}=E_{X_2}-E_{X_0}$ and $\delta_{X_1}=E_{X_2}-E_{X_1}$] exhibit linear and parabolic variation with $V_g$ (reproduced in sample D4 as well, see \textbf{Supplementary Figure 8}), in excellent agreement with the above analysis. We note that such Stark effect is unconventional since the usual quantum confined Stark  effect (QCSE) in quantum wells, where the applied vertical electric field is uniform, results in a perturbing potential having odd parity. Thus the first-order QCSE (linear) is usually negligible, and we only observe a parabolic shift in the emission energy due to the second-order correction \cite{singh2007electronic, abraham2021anomalous,das2020highly,verzhbitskiy2019suppressed,klein2016stark}.

\textbf{Gate tunable exciton lifetime:} Figure \ref{fig:Gate dependent trpl}a shows the peak-resolved (spectral resolution of 0.8 meV) TRPL spectra (see \textbf{Methods}) for $X_0$, $X_1$, and $X_2$, at $V_g=0$ and $3$ V, suggesting a faster decay at higher $V_g$ for all the ILE peaks. The transient response is captured well (solid black lines in Figure \ref{fig:Gate dependent trpl}a) by a set of rate equations and Gaussian formation model (see \textbf{Methods}, equations \ref{diff eqn_0}-\ref{diff eqn_2}). The extracted decay ($\tau_i$) and formation time ($\tau_{fi}$) are plotted for the exciton $X_i$, $i=0,1,2$ in Figure \ref{fig:Gate dependent trpl}b-c. Around $V_g=0$ V, the decay time varies over 10-fold from $X_0$ ($\sim 100$ ns) to $X_2$ ($\sim 9$ ns). However, at large $V_g$, all the three ILEs show similar decay time (4-6 ns). On the other hand, the formation times are relatively weaker function of $V_g$ and reduces slightly with increasing $V_g$.

The kinetics can be understood by the cascaded processes schematically depicted in Figure \ref{fig:Gate dependent trpl}d. At small $V_g$, the respective net lifetimes follow the trend $\tau_0 \gg \tau_1 > \tau_2$ (Figure \ref{fig:Gate dependent trpl}b), which is readily understood due to the additional non-radiative decay paths $\gamma_{20}$ and $\gamma_{21}$ for $X_2$, and $\gamma_{10}$ for $X_1$. The order of the respective formation times ($\tau_{f0}= 5.6$ ns, $\tau_{f1}= 3.6$ ns, and $\tau_{f2}= 1.1$ ns) in Figure \ref{fig:Gate dependent trpl}c, also supports the model of cascaded formation. In addition, a longer lifetime would mean the state is blocked for a longer duration, increasing the formation time.

The strong gate dependence of the ILE lifetime is captured through a simple model where the gate dependent non-radiative process is considered as proportional to induced carrier density (see equations \ref{decay time scales}-\ref{eq:decay time scales2} in \textbf{Methods}):
\begin{equation}\label{eq:lifetime_model}
  \tau_i(V_g)= \Bigg{[}\frac{1}{\tau_i(V_g=0)} + C_i(e^{\alpha V_g} - 1)\Bigg{]}^{-1}
\end{equation}
The model (solid traces in Figure \ref{fig:Gate dependent trpl}b) accurately reproduces the $V_g$ dependent lifetime values (symbols) by using $\alpha$ and $C_i$ as fitting parameters. We observe a $V_g$-modulation of $\tau_0$ by more than 20-fold from $100$ to $5$ ns (Figure \ref{fig:Gate dependent trpl}b), which correlates well with the PL intensity reduction of $X_0$ with $V_g$, in Figure \ref{fig:gate dependent ILE}a. This is a direct evidence of the gate-induced non-radiative process due to the delocalization of the electron in the flattened conduction band (Figure \ref{fig:gate dependent ILE}d). $X_0$ being the ground state of the well, the inter-excitonic transfer related non-radiative decay channels (Figure \ref{fig:Gate dependent trpl}d) are suppressed. On the other hand, At low $V_g$, $\tau_1$ and $\tau_2$ are dominated by the (gate independent) non-radiative decay channels to other lower energy states (that is, $\gamma_{10}$, $\gamma_{20}$, and $\gamma_{21}$), hence remain nearly unchanged up to $V_g=2$ V (Figure \ref{fig:Gate dependent trpl}b). The $V_g$-dependent non-radiative decay rate starts dominating only at large $V_g$ for $X_1$ and $X_2$, resulting in a reduction of $\tau_{1}$ and $\tau_{2}$.

\textbf{Optical power induced anharmonicity:} We now vary $P$ over nearly two decades using a pulsed laser (531 nm) at $V_g=0$ V and plot the ILE peak positions in Figure \ref{fig:power dependent trpl}a. While $X_0$ exhibits a strong blueshift ($\approx 1$ meV$/\mu$W), the shift for $X_1$ and $X_2$ is negligible. Hence, the inter-excitonic separations ($\delta E_{21}$ and $\delta E_{10}$) do not remain equal at higher $P$, suggesting departure from harmonic behaviour. Such anharmonicity and power-dependent blueshift can be understood by the perturbing potential ($U_{dd}$) arising from ILE dipolar repulsion \cite{sun2022excitonic,laikhtman2009exciton}:
\begin{equation} \label{eq:blueshift}
     U_{dd} = \int nU(r)d^2r = n\frac{q^2d}{\epsilon_0\epsilon_r}
 \end{equation}
where $U(r) = \frac{q^2}{2\pi\epsilon_0\epsilon_r}(\frac{1}{r} - \frac{1}{\sqrt{r^2 + d^2}})$ is the repulsion between two ILE dipoles placed at a distance r (schematically shown in Figure \ref{fig:power dependent trpl}b, left panel). $\epsilon_0$ is the vacuum permittivity, $\epsilon_r$ is the effective relative permittivity of the heterojunction, $n$ is the effective concentration of exciton dipoles, and $d$ is the interlayer separation. Due to this induced anharmonicity, it is expected to observe a lifting of degeneracy for $X_1$ and $X_2$, as shown schematically in Figure \ref{fig:power dependent trpl}b (right panel). Since the lifetime of $X_0$ is significantly larger than that of $X_1$ and $X_2$, the steady-state density (generation rate $\times$ lifetime) of ILE dipoles is dominated by the population of $X_0$ ($n_{X0}$). Since $I_{X0} (\propto n_{X_0})  \propto P^{0.34}$ (see Figure \ref{fig:introduction to three moire ILEs}h), equation \ref{eq:blueshift} indicates that the blueshift ($E_{dd}$) of $X_0$ should follow $E_{dd} \propto P^{0.34}$, in good agreement with the linear fit in Figure \ref{fig:power dependent trpl}c. From equation \ref{eq:blueshift}, $n_{X0}$ is estimated to be $\approx 9.5\times 10^{11}$ cm$^{-2}$ (which is less than $n_0/2$) at the highest optical power used ($17.7$ $\mu$W).

To the best of our knowledge, the observed average rate of the blueshift with power for $X_0$ ($\approx$ 1 meV/$\mu$W) is the highest reported value for ILE to date \cite{sun2022excitonic,nagler2017interlayer,rivera2015observation,unuchek2019valley}, indicating a strong inter-excitonic interaction. The strong confinement of $X_0$ does not allow it to drift out of the moir\'e trap in the presence of such dipole-dipole repulsion, resulting in a large blueshift. On the other hand, weaker confinement of ${X_1}$ and ${X_2}$ allows them to drift away under such dipolar repulsion, resulting in a suppressed blueshift in this small power regime.

Figure \ref{fig:power dependent trpl}d, top panel (open symbols) shows the optical power dependent lifetime of $X_0$, $X_1$, and $X_2$. We notice that the lifetime for all the three species is a weak function of $P$. This is in stark contrast with intra-layer free exciton where Auger effect drastically reduces the lifetime at higher $P$ \cite{kumar2014exciton,kuroda2020dark}. Such a weak dependence of lifetime on $P$ is a result of protection from Auger-induced exciton-exciton annihilation due to a combined effect of moir\'e trapping and strong dipolar repulsion.

For a perfect two-dimensional harmonic well, $X_0$, $X_1$, and $X_2$ are expected to exhibit a degeneracy of 1, 2, and 3, respectively. Through the optically induced anharmonicity, we expect the degeneracy of $X_1$ and $X_2$ to be lifted (Figure \ref{fig:power dependent trpl}b, right panel). However, our simulation suggests only $< 2$ meV fine-splitting, and the inhomogeneous broadening of the peaks does not allow us to observe such small splitting in the emission spectra.

Interestingly, while $X_2$ exhibits a mono-exponential decay at low power, its dynamics becomes bi-exponential at higher power ($P > 3.9$ $\mu$W) with an additional lifetime of $\tau_a \sim 1$ ns, as indicated by the blue solid symbols in Figure \ref{fig:power dependent trpl}d (top panel), and the TRPL spectra in the top panels of Figure \ref{fig:power dependent trpl}e-f. In the bottom panel of Figure \ref{fig:power dependent trpl}d, we quantify the degree of anharmonic perturbation by plotting, from Figure \ref{fig:power dependent trpl}a, the relative magnitude of the peak separation ($\delta E = \frac{\delta E_{21} - \delta E_{10}}{\delta E_{21}}\times100\%$) with incident power (0\% corresponding to the harmonic case). The strong correlation between the appearance of the faster additional decay (in region 2) and the strength of the anharmonic perturbation is evident. The faster additional decay likely arises from the fine-split higher energy state of $X_2$, which has reduced confinement into the moir\'e trap, thus having enhanced decay rate (schematically shown in Figure \ref{fig:power dependent trpl}b, right panel). Note that the decay of $X_0$ remains mono-exponential even at higher power since the ground state is non-degenerate (bottom panels of Figure \ref{fig:power dependent trpl}e-f).

In summary, we have shown that the exciton moir\'e potential in heterobilayer can be dynamically tuned through external stimuli, such as gate voltage and optical power. The usual harmonic approximation of moir\'e potential breaks down under such perturbation. The strength of such tunability is evidenced through moir\'e excitons exhibiting (a) confinement dependent tuning of features, (b) anomalous Stark shift where parity is reversed with respect to conventional quantum-confined Stark effect, (c) strong modulation of the lifetime and the inter-excitonic separation, and (d) a giant spectral blueshift through dipolar repulsion. The results will lead to intriguing experiments and applications exploiting dynamic tuning of moir\'e potential.

\section*{Methods}
\textbf{Device fabrication:} We prepared the hBN capped WS\tsub2/WSe\tsub2 heterojunctions using a sequential dry-transfer method (with micromanipulators) where the individual layers were exfoliated from flux grown crystals (HQ-Graphene) on polydimethylsiloxane (PDMS) using Scotch tape. For back gated samples, the pre-patterned metal electrodes are prepared using photolithography followed by sputtering of Ni/Au (10/50 nm) and lift-off. The entire stack (for D1 and D4) is gated from the backside (from the WS$_2$ side) through hBN layer (dielectric) and the pre-patterned metal line. The WS\tsub2 layer is contacted to a different electrode (Gr) for carrier injection. After completion of the transfer process, the devices are annealed inside a vacuum chamber ($10^{-6}$ mbar) at 250$^\circ$C for 5 hours for better adhesion of the layers and removal of air bubbles. The angle and stacking between WS\tsub2/WSe\tsub2 layers are confirmed using SHG (see \textbf{Supplementary Figure 1}).

\textbf{PL measurement:} All the PL measurements on the samples are carried out in a closed-cycle cryostat at 4.5 K using a $\times$50 objective (0.5 numerical aperture) lense. The bottom gate voltages are applied using a Keithley 2636B source meter (for both PL and TRPL), and then the PL spectra are collected using a spectrometer with 1800 lines per mm grating and CCD (Renishaw spectrometer). We use the 532 nm CW and 531 nm pulsed lasers to excite the sample. The spot size for both pulsed and CW laser is $\sim$1.5 $\mu$m. All the power values are measured using a silicon photodetector from Edmund Optics. All the error bars in different plots in the manuscript indicate mean $\pm$ standard deviation.

\textbf{TRPL measurement:} Our custom-built TRPL setup comprises of a 531 nm pulsed laser head (LDH-D-TA-530B from PicoQuant) controlled by the PDL-800 D driver, a photon-counting detector (SPD-050-CTC from Micro Photon Devices), and a time-correlated single photon counting (TCSPC) system (PicoHarp 300 from PicoQuant). The pulse width of the laser is 40 ps. For the spectrally resolved TRPL from moir\'e ILEs, a combination of a long pass filter (cut in wavelength of 650 nm) and a wavelength-tunable monochromator (Edmund optics, 2 cm$^2$ Square holographic gratings) with 0.5 nm resolution (corresponding to about 0.8 meV resolution in the ILE spectral regime) are placed in front of the SPD. The peak position of the emission from ILEs are simultaneously measured along with TRPL measurement by performing in-situ PL (see Supplemental Material in ref. \cite{chatterjee2022trion} for setup schematic). The instrument response function (IRF) has a full-width-at-half-maximum (fwhm) of $52$ ps.

\textbf{Exciton formation and decay model:} To fit the experimentally obtained TRPL data, we use three differential equations:
\begin{equation}\label{diff eqn_0}
    \frac{dn_0(t)}{dt} = f_0(t) - \frac{n_0(t)}{\tau_0}
    \end{equation}
    \begin{equation}\label{diff eqn_1}
    \frac{dn_1(t)}{dt} = f_1(t) - \frac{n_1(t)}{\tau_1}
    \end{equation}
      \begin{equation}\label{diff eqn_2}
    \frac{dn_2(t)}{dt} = f_2(t) - \frac{n_2(t)}{\tau_2}
\end{equation}
Here $n_i(t)$ is the time dependent population density, $\tau_i$ is the net decay time, and $f_i(t) = \frac{1}{\sigma_i \sqrt{2\pi}} e^{\frac{-(t-\tau_{fi})^2}{2\sigma_i^2}}$ is the Gaussian formation function, and $\tau_{fi}$ is the formation time measured from the laser excitation time for exciton $X_i$, $i=0,1,2$. After solving these equations numerically, we fit the measured TRPL data from the three moir\'e exciton emissions using $\tau_{fi}$, $\sigma_i$, and $\tau_i$ as fitting parameter.

\textbf{Model for gate-voltage dependent lifetime:} The net decay time ($\tau_i$) measured in TRPL (Figure \ref{fig:Gate dependent trpl}b), for exciton $X_i$ ($i=0, 1, 2$) is given by:
\begin{equation}\label{decay time scales}
  \frac{1}{\tau_i(V_g)}= \frac{1}{\tau_{r,i}} + \frac{1}{\tau_{nr0,i}} + \frac{1}{\tau_{nrg,i}(V_g)}
\end{equation}
where $\tau_{r,i}$, $\tau_{nr0,i}$, and $\tau_{nrg,i}(V_g)$ represent the radiative lifetime, gate voltage independent non-radiative lifetime, and the gate voltage dependent non-radiative lifetime, respectively. From Figure \ref{fig:Gate dependent trpl}d, $\frac{1}{\tau_{nr0,2}}= \gamma_{20} + \gamma_{21} + \gamma^\prime_{2}$ for $X_2$, and $\frac{1}{\tau_{nr0,1}}= \gamma_{10} + \gamma^\prime_{1}$ for $X_1$, and $\frac{1}{\tau_{nr0,0}}= \gamma^\prime_{0}$, where $\gamma^\prime_{i}$ is the rate of any other unaccounted non-radiative process for exciton $X_i$. Considering that the rate of the gate dependent non-radiative process is proportional to induced carrier density, which in turn is an exponential function of $V_g$, we write $\frac{1}{\tau_{nrg,i}}= C_ie^{\alpha V_g}$, where $C_i$ and $\alpha$ are fitting parameters. By noting that $\frac{1}{\tau_{r,i}}$ is relative small (in equation \ref{decay time scales}) and becomes smaller with an increase in $V_g$, we write
\begin{equation}\label{eq:decay time scales2}
  \frac{1}{\tau_i(V_g)} \approx \frac{1}{\tau_i(V_g=0)} + C_i(e^{\alpha V_g} - 1)
\end{equation}
\section*{Data Availability}
The data that support the findings of this study are available within the main text and Supplementary Information. Any other relevant data are available from the corresponding authors upon request.

\section*{Acknowledgements}
S.C. and K.M. acknowledge useful discussions with Garima Gupta, Nithin Abraham, Mayank Chhaperwal, and Manish Jain. K.W. and T.T. acknowledge support from the JSPS KAKENHI (Grant Numbers 19H05790 and 20H00354). K.M. acknowledges the support from a grant from Science and Engineering Research Board (SERB) under Core Research Grant, a grant from the Indian Space Research Organization (ISRO), a grant from MHRD under STARS, and support from MHRD, MeitY, and DST Nano Mission through NNetRA.
\section*{Author contribution}
K.M. designed the experiment. M.D., S.C., and S.D. fabricated the devices and conducted the measurements. P.D. conducted the electrostatic simulation. R.B. and V.R. performed the SHG measurements for all samples. K.W. and T.T. grew the hBN crystals. S.C., M.D., and K.M. conducted the data analysis and wrote the manuscript with inputs from others.
\section*{Competing Interests}
The authors declare no competing interests.
\newpage
\newpage
\begin{figure}[!hbt]
	\centering
	\vs{-0.1in}
	\hs{-0in}
   \includegraphics[scale=0.46]{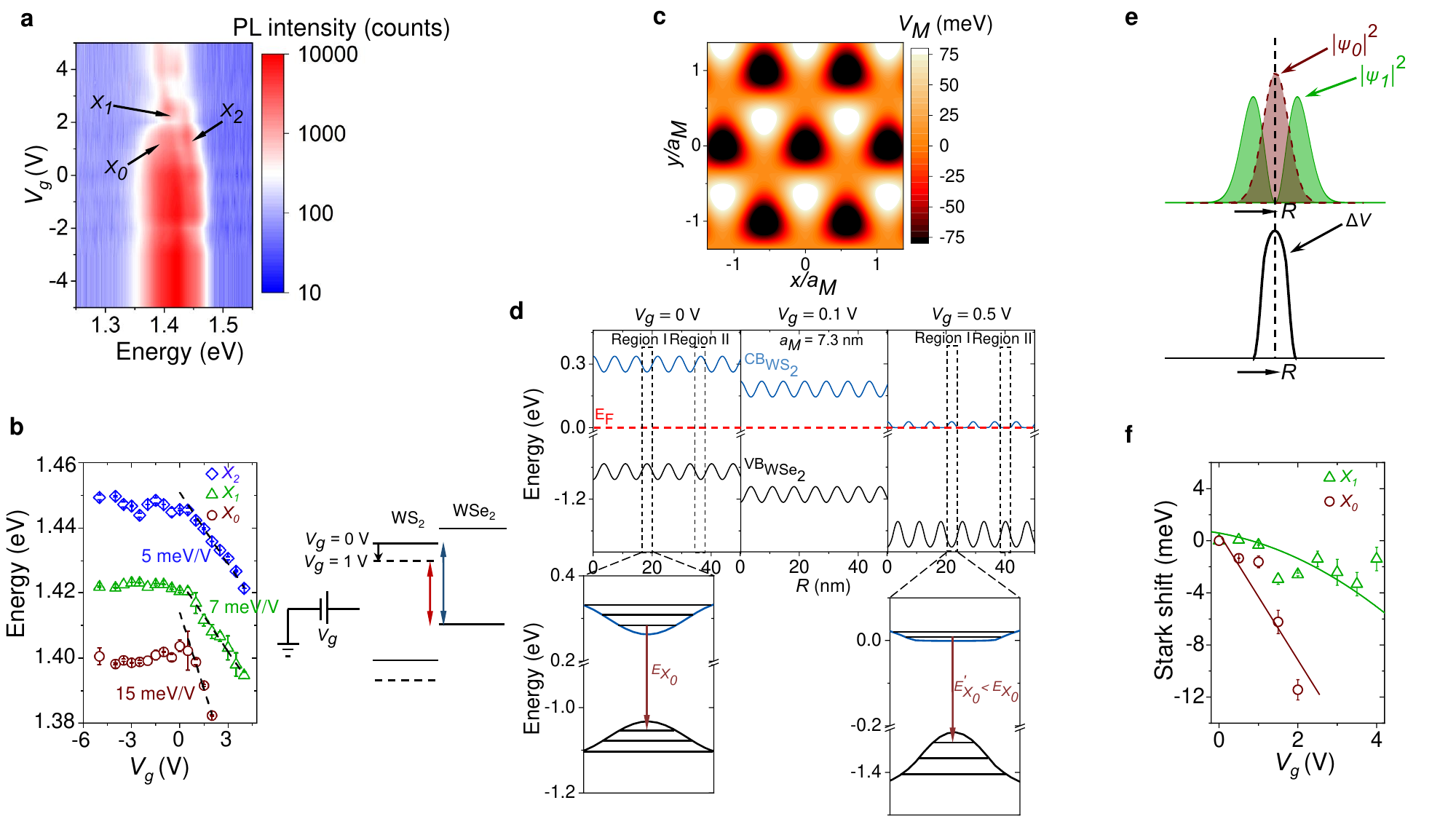}
	\vspace{-0.1in}
	\caption{\textbf{Gate-tunable moir\'e potential and unconventional Stark effect.} (a) Color plot of $V_g$ dependent PL spectra showing $X_0$, $X_1$ and $X_2$ resonances. (b) Left panel: Fitted peak positions showing the gradual redshift of the three ILE peaks with $V_g$. The black dashed lines indicate guide-to-eye in the $V_g>0$ regime. Right panel: Interlayer bandgap reduction is shown schematically with increasing $V_g$. (c) 2D projection of the variation of the calculated moir\'e potential. (d) Top panel: Simulated conduction and valance band profile at three different $V_g$ ($0$, $0.1$, and $0.5$ V) values obtained by solving the 1D Poisson equation with the moir\'e potential fluctuation (see Supplementary Note 3 for details). For simulation, the thickness of the gate dielectric (hBN) is assumed to be 20 nm. Region I (II) in the top left panel denotes the minimum (maximum) energy of the WS\tsub2 conduction band due to moir\'e potential induced spatial energy fluctuation. At lower $V_g$ (top middle panel), the conduction band gradually comes down in energy towards the Fermi level (red dashed line) maintaining the same degree of fluctuation. At higher $V_g$ (top right panel), when the conduction band is close to the Fermi level, it starts flattening due to screening. This also results in a deepening in the valence band fluctuation. Bottom panel: Zoomed-in Region I at $V_g=0$ V (in left) and $V_g=0.5$ V (in right). The transition energy for $X_0$ ($E_{X_0}$, shown by arrow) decreases at higher $V_g$. (e) $|\psi_i|^2$  ($i=0,1$) plotted along with the in-plane perturbing potential $\Delta V$ indicating strong overlap (non-overlap) between $\Delta V$ and $|\psi_0|^2$ ($|\psi_1|^2$) due to different parity of the wave functions. (f) Stark shift of $X_0$ ($\delta_{X_0}$) and $X_1$ ($\delta_{X_1}$) plotted with $V_g$. $\delta_{X_0}$ ($\delta_{X_1}$) shows a linear(parabolic) Stark shift fitting (solid traces).} \label{fig:gate dependent ILE}
\end{figure}
\newpage
\begin{figure}[!hbt]
	\centering
	\vs{-0.1in}
	\hs{-0in}
   \includegraphics[scale=0.46]{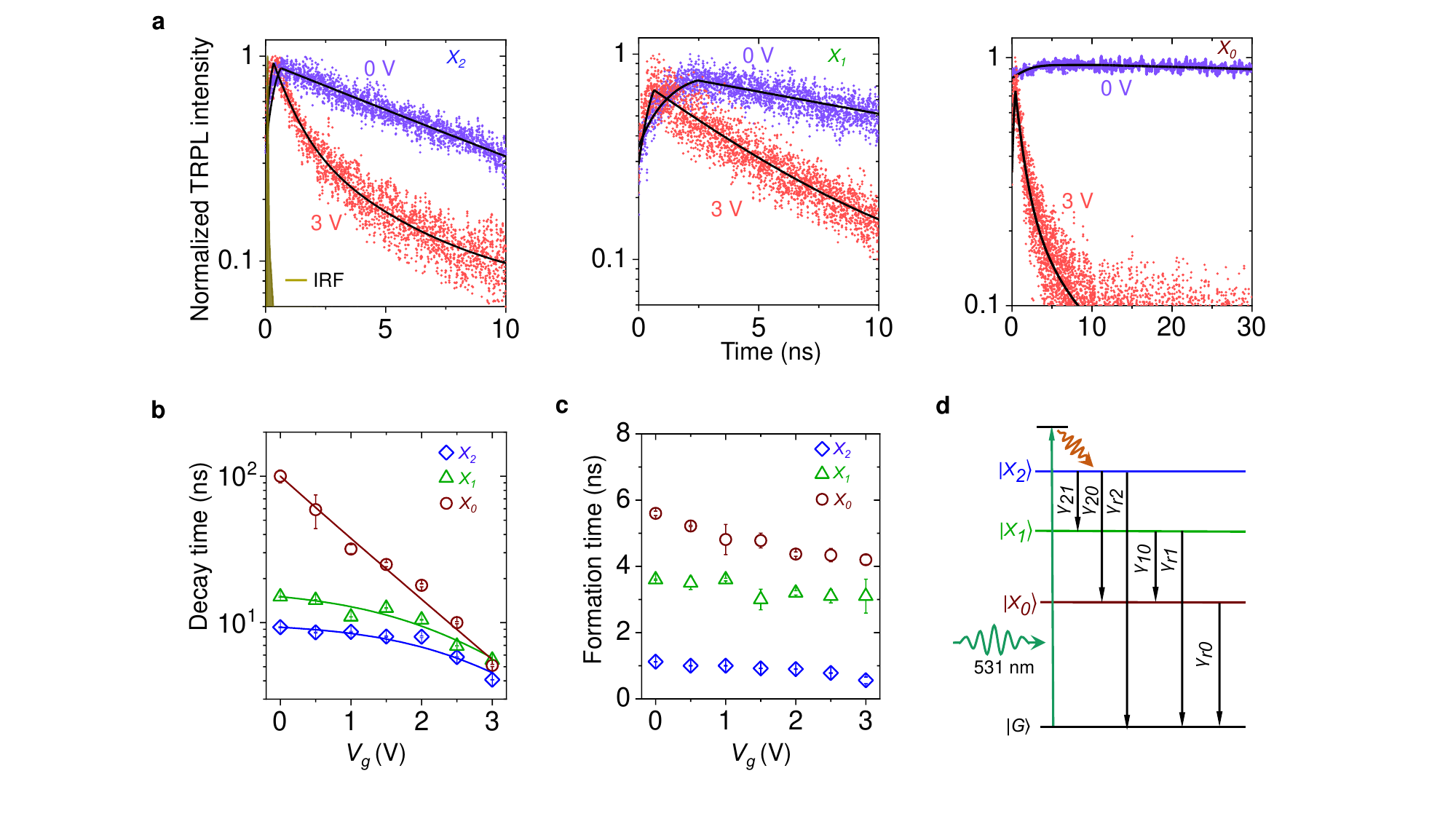}
	\vspace{-0.1in}
	\caption{\textbf{Gate induced lifetime modulation of moir\'e exciton.} (a) Peak resolved TRPL spectra (symbols) along with model (described in \textbf{Methods}) predicted fitting (black trace) at $V_g=0$ and 3 V for $X_0$, $X_1$, and $X_2$. The IRF is shown in the left panel. (b) Extracted decay time (symbols) for different moir\'e ILEs as a function of $V_g$. Solid traces represent the model (equation \ref{eq:lifetime_model}) prediction. (c) Extracted formation times plotted as a function of $V_g$. (d) Cascaded formation process for different ILEs, showing radiative ($\gamma_{r,i}$) for the exciton $X_i$ ($i=0, 1, 2$), and inter-excitonic non-radiative paths ($\gamma_{ij}$) between excitons $X_i$ and $X_j$.} \label{fig:Gate dependent trpl}
\end{figure}
\newpage
\begin{figure}[!hbt]
	\centering
	\vs{-0.1in}
	\hs{-0.1in}
	\includegraphics[scale=0.46]{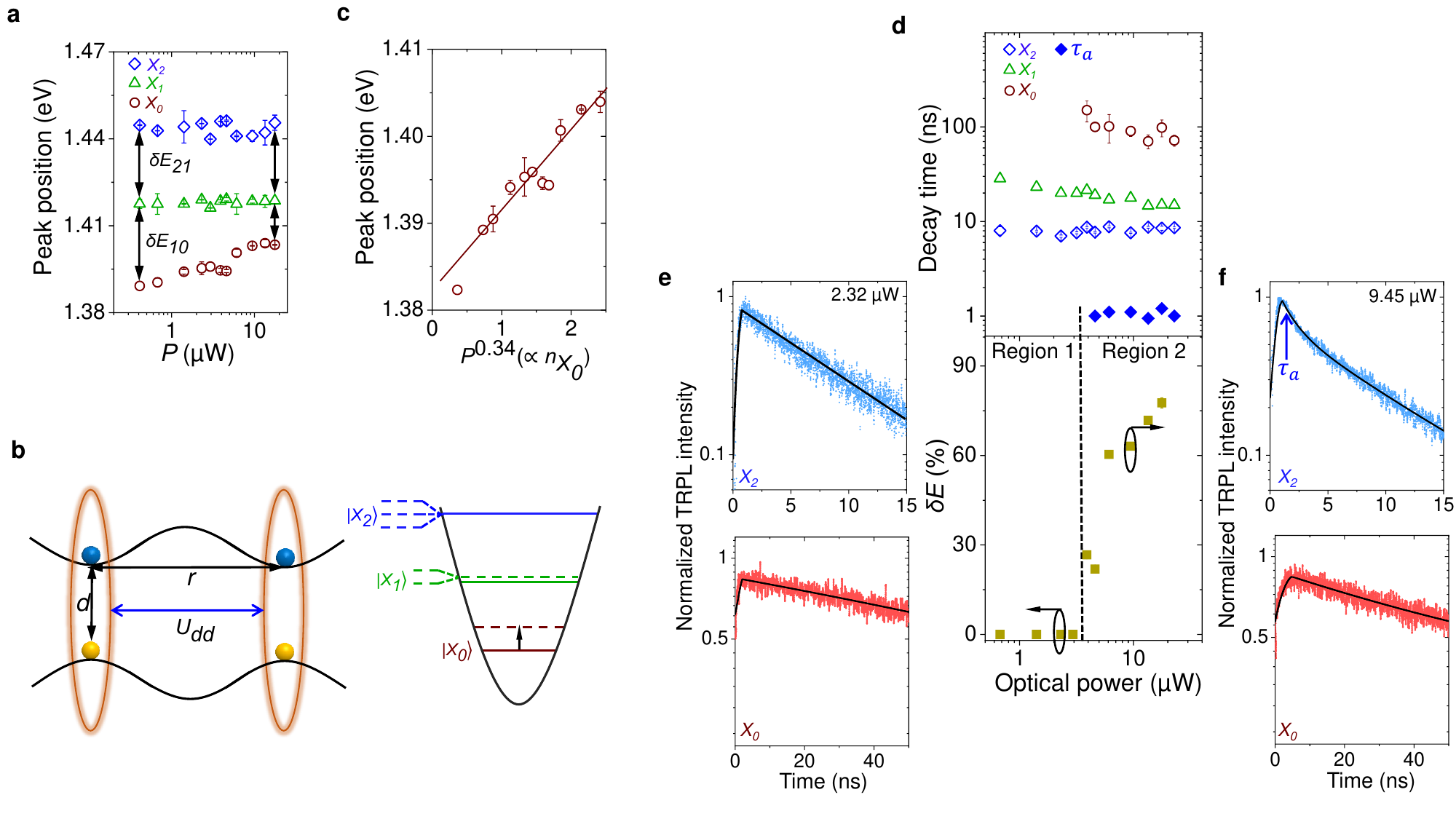}
	\vspace{-0.1in}
	\caption{\textbf{Optical power dependent anharmonic tuning of moir\'e potential.} (a) PL peak position for $X_0$, $X_1$, and $X_2$, plotted against optical power ($P$). $X_0$ exhibits a strong blueshift (1 meV/$\mu$W) with $P$. The inter-excitonic peak separation is similar at low $P$, but becomes different at high $P$. (b) Left panel: Schematic representation of the interlayer excitonic dipole repulsion model. Right panel: Lifting of degeneracy for $X_2$ and $X_1$ in a two-dimensional harmonic oscillator shown schematically at higher $P$. Dipole repulsion results in blueshift of the states (dotted line), which is highest for $X_0$ (shown by a black arrow). (c) Peak position of $X_0$ (symbols) plotted against $P^{0.34} (\propto n_{X0})$, showing excellent linear fit. (d) Top panel: Extracted lifetime of $X_0$, $X_1$, and $X_2$ (in open symbols) plotted with optical power, showing a weak dependence due to suppressed Auger process. The solid blue symbols ($\tau_a$) indicate additional decay path of $X_2$ due to anharmonicity induced degeneracy lifting at higher $P$. Bottom panel: Percentage change in the inter-exciton peak separation with $P$, indicating the degree of anharmonicity induced by $P$. The Regions 1 (harmonic) and 2 (anharmonic) are separated by a dashed black line, and correlates well with the appearance of $\tau_a$ in $X_2$. (e-f) The top and bottom panels show the TRPL spectra for $X_0$ and $X_2$, at (e) $P=2.32$ and (f) 9.45 $\mu$W, respectively. $X_2$ decay becomes bi-exponential with a fast ($\approx$ 1 ns) $\tau_a$ at higher $P$, while $X_0$ decay remains mono-exponential all through.} \label{fig:power dependent trpl}
 \end{figure}
\AtEndDocument{

\section*{Supplementary material}

# Supplementary Note 1:

**Polarization dependent SHG measurement technique:** A SHG microscopy setup is used to measure the twist angle between the $WS_2$ and $WSe_2$ layers. The laser setup consists of a fixed-wavelength femtosecond pulsed laser source (Fidelity HP) centered at 1040 nm with a pulse width of 140 fs and a repetition rate of 80 MHz, which pumps an optical parametric oscillator (Levante-IR) to generate a signal source at 1600 nm wavelength. The signal is fed to a Harmonic generator (HarmoniXX - APE) to produce a harmonic wavelength at 800 nm, which serves as the input laser source for the polarization dependent SHG measurement.

The hetero-bilayer sample is placed on a custom-built rotational mount in an inverted position on the x-y stage of an inverted optical microscope (Olympus IX73). The 800 nm harmonic input laser source is then focused onto the sample using a $\times$ 20 (numerical aperture of 0.75) objective, and the backward second harmonic signal centred at 400 nm was collected using the same objective. The emitted SHG signal is then spatially separated from the input laser source using a dichroic filter and detected using a photomultiplier tube. Bandpass (FWHM of 40 nm) and short-pass filters (cut in wavelength 550 nm) are mounted before the photomultiplier tube to reject any residual pump source and minimize background signal.

For twist angle SHG measurement, the sample is rotated from 0° to 180° using the rotational mount. For each rotation, a pair of galvo mirrors are used to scan the input laser source over the flake point-by-point in a raster pattern over a field of view of 100 $\mu$m $\times$ 100 $\mu$m to acquire a two-dimensional SHG intensity map. A 256 $\times$ 256 pixels SHG image is obtained by mapping the PMT signal in every pixel with a resolution of $\approx$ 396 nm. An analyzer is placed in front of the PMT to selectively detect the horizontal component of the SHG signal from the sample with respect to the laboratory horizontal axis.

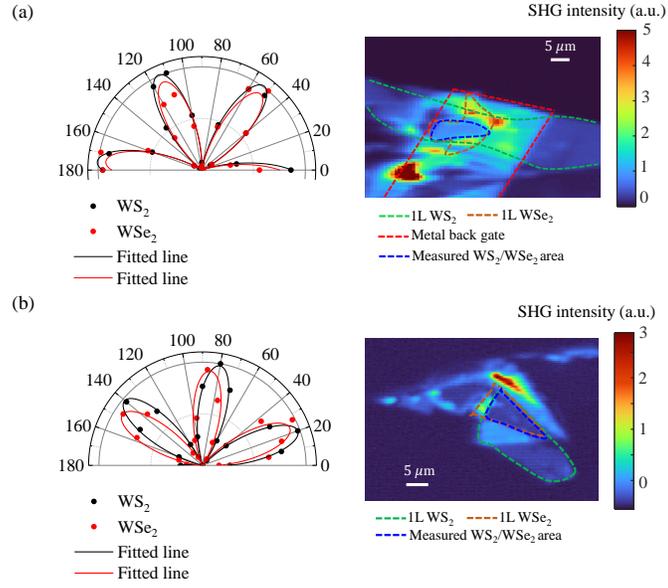

Figure 1: **Polarization-resolved SHG result from the twisted WS$_2$/WSe$_2$ stacks.** (a) Polarization resolved SHG polar plot from isolated WS$_2$ and WSe$_2$ monolayers is represented for sample D1 in the left panel. Right panel shows SHG map of the full sample (D1), where the isolated 1L WS$_2$ (green), WSe$_2$ (orange), and the measured heterojuction area (blue) are marked by dashed lines. From left panel the twist angle is confirmed as $\sim 1°$. The junction area in the the SHG color plot (right panel) shows a drop in intensity, confirming the twist is around AB stacking (60°). (b) Same as (a), but for sample D2 confirming the twist angle to be $\sim 6°$ around AB stacking.

3

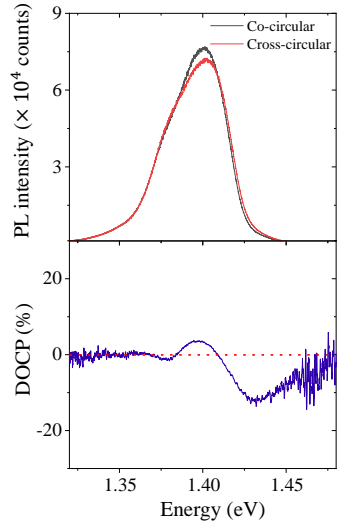

Figure 2: **Circular polarization resolved interlayer excitonic emission.** Circular polarization resolved interlayer excitonic emission for sample D1, with 705 nm (close to WSe$_2$ intralyer excitonic resonance) excitation. The top and bottom panels show the circular polarization resolved spectra and the degree of circular polarization (DOCP). The alternating sign of the DOCP for different ILE peaks indicates their moiré excitonic nature.

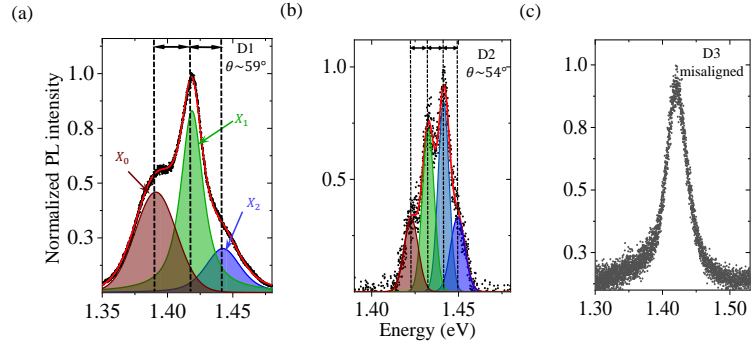

Figure 3: **Twist angle dependence of the moiré exciton emission spectra.** In (a), (b), and (c), the interlayer excitonic PL emission from D1, D2 and D3 is plotted with 532 nm excitation at 4 K. The separation between different ILE peaks reduces from $\sim 25$ meV (in D1 with a twist angle of 59°) to $\sim 9$ meV (in D2 with a twist angle of 54°). The largely misaligned sample D3 shows a single broad peak.

## Supplementary Note 2:

**Moiré potential as two-dimensional harmonic potential well:** The depth of the moiré potential is formulated from the spectral separation of the interlayer excitonic emission peaks (Figure 1e, main text). The emission energies are at $\approx 25$ meV separation from each other (1.392, 1.418 and 1.442 eV for $X_0$, $X_1$, and $X_2$ respectively). This near equal separation suggests an assumption of 2D harmonic oscillator well for moiré pockets is valid [1–4], with energy separation of $\hbar\omega$ (25 meV) between each state. The ground state, $X_0$, must be then at an energy $\hbar\omega$ above the bottom of the well, and this gives rise to a total well depth of $3\hbar\omega = 75$ meV.

The Hamiltonian for interlayer excitons in $WS_2/WSe_2$ can be written as [1,3]:

$$H = \hbar\omega' + \frac{\hbar^2 Q^2}{2M} + \Delta(r) \tag{1}$$

Where $\hbar\omega'$ and $\frac{\hbar^2 Q^2}{2m}$ are the constant and kinetic energy of ILE in the center of mass momenta space. $M$ is the effective mass. Here, $\Delta(r)$ is the periodic moiré potential in real space and it can be projected in two dimensions as: $\Delta(r) = \sum_{n=1}^{6} V_n e^{i\vec{b_n}\cdot\vec{r}}$, where $V_n$ and $\vec{b_n}$ are moiré potential and reciprocal lattice vectors, respectively. $\vec{b_n}$ vectors are calculated from the monolayer lattice vectors, considering 4% lattice mismatch. The 75 meV well depth corresponds to a $V_n$ of $\approx 21$ meV. Thus $\Delta(r)$ is obtained and plotted in Figure 2c (main text).

## Supplementary Note 3:

**Formulation of 1D Poisson equation:** To understand the electrostatics of the $WS_2/WSe_2$ moiré heterojunction, we numerically solve a 2D Poisson equation $[\nabla^2\phi = \sigma\delta(z)]$ where $WS_2/WSe_2$ heterobilayer is considered to be a 2D sheet of charge at $z = 0$ plane with a 2D charge density $\sigma = q(p - n + N)$. Here $p, n$ are the 2D density of electron and hole, respectively. $N$ refers to the unintentional doping in the heterobilayer at zero gate voltage. Under positive gate voltage, $p \approx 0$, and

$$n(x) = N_{2D} log[1 + exp(\frac{E_F - E_C(x) + q\phi(x)}{k_B T})] \qquad (2)$$

where $N_{2D}$ is the 2D density of states. $E_F$ is the Fermi level and is referenced as zero. Both conduction (CB) and valence band (VB) profiles are spatially varying according to the spatial bandgap modulation in a moiré superlattice. Only CB ($WS_2$) and VB ($WSe_2$) profiles at $K$ point $[E_C^K(x), E_V^K(x)]$ corresponding to the direct bandgap are considered. The moiré potential for electrons and holes is assumed to be a simple cosine profile as: $E_C^K(x) = E_{C0}^K(x) + \Delta E_G^K cos(k_x x)$ and $E_V^K(x) = E_{V0}^K(x) - \Delta E_G^K cos(k_x x)$ with a spatially varying bandgap of $E_G^K = (E_{C0}^K - E_{V0}^K) + 2\Delta E_G^K cos(k_x x)$. Here $k_x = 2\pi/a_M$, $a_M = 7.3$ nm is the moiré period, and $\Delta E_G^K = 75$ meV.

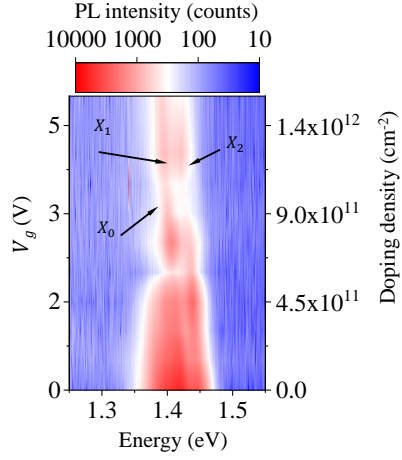

Figure 4: **Gate dependent color plot showing doping density.** With increasing $V_g$, the n-doping density numbers are shown in the right axis. The highest doping density achieved is $1.5 \times 10^{12}$ cm$^{-2}$ at $V_g = 5$V. The doping density is estimated assuming charge neutral voltage as 0 V, and by only taking into account the geometrical capacitance of the hBN dielectric, ignoring any channel capacitance effect which comes in series with the gate dielectric capacitance. Hence the quoted doping density is likely overestimated.

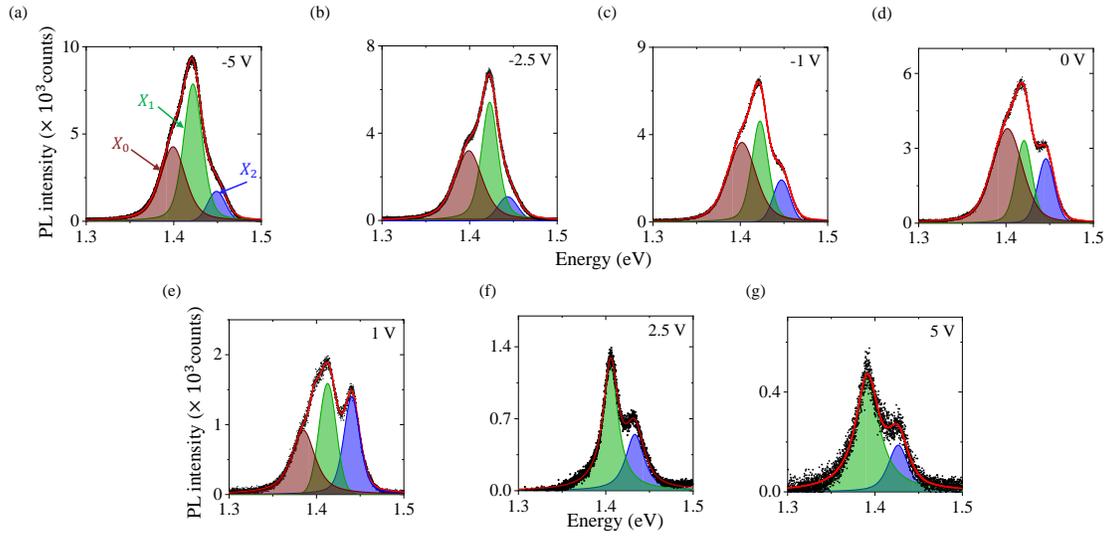

Figure 5: **PL spectra for different gate voltage points.** (a) to (g) represent PL spectra at -5V, -2.5V, -1V, 0V, 1V, 2.5V, and 5V. The three ILE resonances are marked in (a). In the n-doping regime, they evolve gradually, showing quick disappearance of $X_0$ resonance (after $V_g > 2$V).

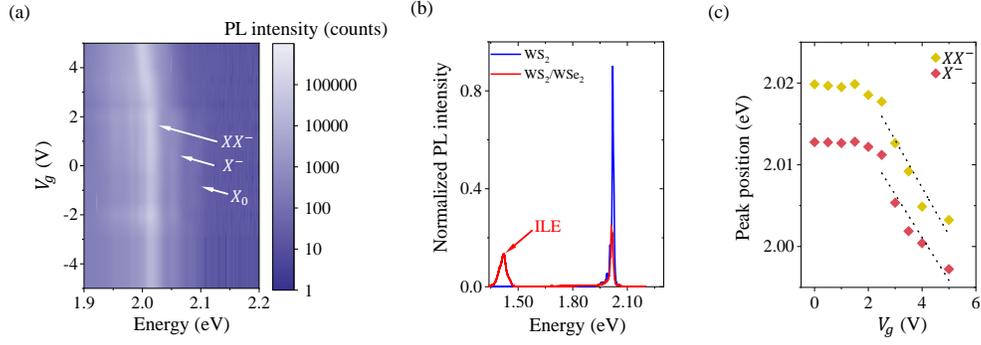

Figure 6: **Trion and biexciton peaks of WS$_2$ with increasing n-doping.** (a) Color plot of both $X^-$ and $XX^-$ of WS$_2$ (from the hetrojunction) showing similar redshift as the ILE peaks. (b) A comparison of PL between isolated 1L WS$_2$ and WS$_2$/WSe$_2$ junction is shown. A clear reduction in WS$_2$ intralayer excitonic emission can be seen due to charge transfer (around 2.02 eV). WSe$_2$ intralayer excitonic emission is almost completely quenched at the heterojunction. This proves strong interlayer coupling. (c) $X^-$ and $XX^-$ of WS$_2$ showing similar redshift ($\approx$5 meV/V, dotted black line) as PL peak shift of $X_2$ ILE with increasing n-doping.

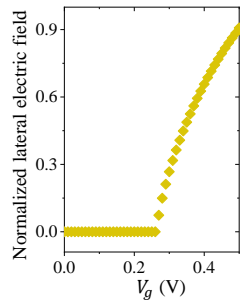

Figure 7: **Local lateral field from Poisson equation.** The calculated local lateral electric field obtained from the Poisson equation as a function of $V_g$.

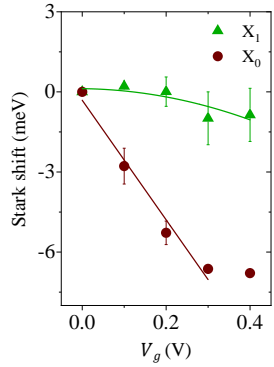

Figure 8: **Stark effect measurement from a different sample.** The Stark effect measurement is repeated in another sample (D4). $X_0$ and $X_1$ exhibit linear and parabolic Stark shift, respectively as a function of the gate voltage. This is in good agreement with the results obtained from sample D1 (Figure 2f of main text).

}
\end{document}